\documentclass[12pt,fleqn,twoside]{article}
\usepackage{espcrc1,epsfig,feynmp}

\usepackage{graphicx}
\usepackage[figuresright]{rotating}

\newcommand{\AmS}{{\protect\the\textfont2
  A\kern-.1667em\lower.5ex\hbox{M}\kern-.125emS}}
\def\nn{\nonumber}

\def\be{\begin{equation}}
\def\ee{\end{equation}}
\def\bea{\begin{eqnarray}}
\def\eea{\end{eqnarray}}

\def\mev{\,{\rm MeV}}
\def\gev{\,{\rm GeV}}

\newcommand{\vk}{{\bf k}_\perp}

\newcommand{\da}{{distribution amplitude}}
\newcommand{\das}{{distribution amplitudes}}
\newcommand{\wf}{wave function}
\newcommand{\wfs}{wave functions}

\renewcommand{\d}{\rm d}
\renewcommand{\b}{\rm b}
\renewcommand{\u}{\rm u}

\newcommand{\bbar}{\overline{\rm b}}
\newcommand{\ubar}{\overline{\rm u}}
\newcommand{\dbar}{\overline{\rm d}}

\newcommand{\gsim}{\raisebox{-3pt}{$\,\stackrel{\textstyle >}{\sim}\,$}}

\title{THE $\b - \u$ SKEWED PARTON DISTRIBUTIONS}

\author{P. Kroll\address{Fachbereich Physik, Universit\"at Wuppertal\\ 
        Gau\ss strasse 20, D-42097 Wuppertal, Germany}%
        \thanks{ Supported in part by the TMR network ERB 4061 Pl 95 0115
                }}

\begin{document}
\sloppy
\thispagestyle{empty}

\begin{flushright}
WU B 00-05 \\
hep-ph/0003097\\
March 2000 \\[20mm]
\end{flushright}

\begin{center}
{\Large\bf THE $\b - \u$ SKEWED PARTON DISTRIBUTIONS}\\[20mm]

{\Large P.\ Kroll}\\[10mm]
{\it Fachbereich Physik, Universit\"at Wuppertal}\\[1mm]
{\it Gau\ss strasse 20, D-42097 Wuppertal, Germany}\\[1mm]%
{ E-mail: kroll@theorie.physik.uni-wuppertal.de}\\[20mm]
\end{center}

\begin{center}
Contribution to the QNP2000 conference\\[5mm] 
Adelaide (February 2000)
\end{center}
\newpage
\thispagestyle{empty}
\vspace*{10cm}
\pagebreak
\setcounter{page}{1}

\maketitle

\begin{abstract}
The $\b$-$\u$ skewed parton distributions (SPDs) are discussed. The
SPDs allow an unambigous superposition of overlap and resonsance
contributions and their zeroth order moments represent the 
$B\to \pi$ transition form factors. The values of the form factors 
at maximum recoil are found to be $F_+(0)=F_0(0)=0.22\pm 0.05$ in agreement 
with measurements of the $B\to \pi\pi$ branching ratio. 
The branching ratios for the semi-leptonic $B\to\pi$ decays
are evaluated too. 
\end{abstract}

\section{INTRODUCTION}
A good theoretical understanding of the  $B\to\pi$ transition form
factors are of utmost interest. Accurate predictions of these form
factors would permit a determination of the less well-known 
Cabbibo-Kobayashi-Maskawa matrix element $V_{\u\b}$ from experimental
rates of semi-leptonic $B\to\pi$ transitions. The $B\to\pi$ form factors  
also form an important ingredient of the calculation of $B\to\pi\pi$ 
decay rates; the factorization hypothesis relates the form factors at 
maximum recoil to the decay rates. Thus, not surprisingly, the
$B\to\pi$ and other heavy-to-light form factors, attracted the
attention of many theoreticians. In several of these approaches two 
distinct dynamical mechanism are considered which build up the
$B\to\pi$ form factors: The $B\pi$ resonances which control the form 
factors at small recoil and the overlap of the meson wave functions 
that dominates at large recoil. Other mechanisms, like the
perturbative one, provide only small and often negligible
corrections. The crucial questions are how to match these two
prominent contributions at intermediate recoil and how to 
avoid double counting. In a recent  article \cite{feld:00}, on which 
I am going to report here, we proposed to start from skewed parton 
distributions \cite{mue98}. The SPDs allow an unambigous superposition 
of $B\pi$ resonances and the overlap contribution. 
\section{$\b$-$\u$  SKEWED PARTON DISTRIBUTIONS}
To be specific let us consider the semi-leptonic decay
$\bar{B}^0\to \pi^+\ell^-\bar{\nu}_l$. Instead of the usual
form factors, $F_+$ and $F_0$, (see e.g.\ 
\cite{Wirbel:1985ji,Khodjamirian:1998ji}) it is more appropriate here 
to use the definition
\begin{equation}
\langle \pi^+; p' | \bar{\u}(0) \gamma_\mu {\b}(0) | \bar{B}^0; p\rangle =
F^{(1)}(q^2) \, p'_\mu  + F^{(2)}(q^2)\,\left(q_\mu - \frac{q^2} {M_B^2} p_\mu\right)\,,
\label{current}
\end{equation}
where $q=p-p'$ and $M_B$ ($M_\pi$) is the $B$ ($\pi$) mass. 
A convenient frame of reference is a generalized Breit frame in which
the mesons move collinearly. In this frame the momentum transfer is given by
\begin{equation}
q^2 = \zeta M_B^2 \left(1 -  \frac{M_\pi^2}{M_B^2 (1-\zeta)}\right)\,,
\label{q-z}
\end{equation} 
where the so-called skewedness parameter, $\zeta$, is defined by the
ratio of light-cone plus components
\begin{equation}
\zeta = \frac{q^+}{p^+} = 1 - \frac{p'^+}{p^+}\,.
\end{equation}
The skewedness parameter $\zeta$ covers the interval $[0,1-M_\pi/M_B]$ 
in parallel with the variation of the momentum transfer from
zero (the lepton mass is neglected) to $q^2_{\rm max}=
(M_B-M_\pi)^2$ in the physical region of the $B\to\pi$ transitions. 
The pion mass is neglected in the calculation of SPDs and form factors. The
advantage of the generalized Breit frame is that the $B\to\pi$ matrix 
element of the current's plus component is related to the form factor 
$F^{(1)}$ solely while that of the minus component is related to $F^{(2)}$. 
The matrix elements of the transverse currents are zero. 

The flavour non-diagonal $\b$-$\u$ SPDs 
$\tilde{{\cal F}}^{(i)}_{\zeta}$, $i=1,2$ are defined by the $B - \pi$ 
matrix elements of bilocal products of quark field operators, e.g.\ 
\begin{equation} 
\int \frac{{\d}z^-}{2\pi}\, e^{ixp^+z^-}\, \langle \pi^+; p' |
     \bar{\u}(0) \gamma^+ {\b}(z^-)| \bar{B}^0; p \rangle
                     = (1-\zeta)\, \tilde{{\cal F}}^{(1)}_{\zeta}(x,q^2)\,.
\label{spd-def}
\end{equation}
$x=k^+/p^+$ is the usual fraction of plus-components of the
$\b$-quark and $B$-meson momenta. The second SPD, 
$\tilde{{\cal F}}^{(2)}_{\zeta}$, is analogously defined  with
$\gamma^+$ being replaced by $\gamma^-$. The variable $q^2$ is
redundant in the generalized Breit frame, see Eq.\ (\ref{q-z}).
Integration of (\ref{spd-def}) over $x$ reduces the bilocal operator
product to the local one that defines the form factors, see
(\ref{current}). Hence, one has the reduction formula
\begin{equation}
   F^{(i)}(q^2)\,=\, \int^1_{-1+\zeta} {\d}{x} \, \widetilde{{\cal
                                              F}}^{(i)}_{\zeta}(x)
\label{fff}
\end{equation}
for $i=1,2$. 

Depending on the value of $x$, the SPDs describe different physical
situations:\\
i) For $1\geq x\geq \zeta$ a $\b$-quark with momentum fraction $x$
is emitted from the $B$-meson and a ${\u}$-quark carrying a momentum 
fraction $x'=k^{+}{}'/p^{+}{}'$ is absorbed, turning the $B$-meson 
into a pion. This part of the SPDs can be modelled as overlaps of 
$B$ and $\pi$ light-cone \wfs. For the valence Fock 
states, for instance, the overlap reads
\begin{equation}
\widetilde{{\cal F}}_{\zeta\,{\rm ove}}^{(1)}(x)\,=\,
             \frac{2}{1-\zeta} \, \int \frac{{\d}^2 \vk}{16\pi^3}
   \;\Psi_\pi^{*}(x'=\frac{x-\zeta}{1-\zeta}, \vk)\;
               \Psi_B (x,\vk) \,, 
\label{par-spd}
\end{equation}
where $\vk$ is the intrinsic transverse momentum of the $\b$ ($\u$) 
quark with respect to the $B$ ($\pi$)-meson momentum.\\ 
ii) For $0\leq x<\zeta$ the fraction $x'$ is negative. Interpreting
a parton with a negative momentum fraction as an antiparton with a
positive fraction, one sees that the physical situation is now the
emission of a ${\b}\ubar$ pair from $B$-meson and the formation of the
pion from the remaining partons. This contribution can be desribed by 
overlaps for $N+2\to N$ parton processes; it is found to be very small 
numerically. $B\pi$ resonances contribute to the SPDs in that region 
as well \cite{Radyushkin:1998bz}. The contribution of the most
important resonance, the $B^{*-}$ vector meson, reads
\bea
\widetilde{{\cal F}}_{\zeta\, {\rm res}}^{(1)}(x) &=&
                           \frac{f_{B^*} \, g_{B B^*\pi}}{M_{B^*}} \,
                           \left(M_{B^*}^2 - \frac12 \zeta M_B^2 \right) \,
                           \frac{\phi_{B^*}(x/\zeta)}
                           {M_{B^*}^2 - \zeta \, M_B^2} \nn \ , \\[0.3em]
\widetilde{{\cal F}}_{\zeta\, {\rm res}}^{(2)}(x) &=&
                           \qquad - \frac12 \, M_B^2 \,
                           \frac{f_{B^*} \, g_{B B^*\pi}}{M_{B^*}} \,
                           \frac{\phi_{B^*}(x/\zeta)}
                           {M_{B^*}^2 - \zeta \, M_B^2}\,,
\label{eq:pole}
\eea
where $\phi_{B^*}(y)$ is the valence Fock state distribution amplitude
of the $B^*$-meson. The argument of the $B^*$ \da{}, $x/\zeta$, equals
the momentum fraction, $k_+/q_+$, which the $\b$-quarks carries
w.r.t.\ the $B^*$-meson. \\ 
iii) For $-1+\zeta\leq x <0$ where $x'$ is negative too, the SPDs
describe the emission of a $\bbar$-quark and the absorption of a
$\ubar$ one. Since the probability of finding a $\b\bbar$ sea-quark
pair in the $B$-meson is practically zero, 
$\widetilde{{\cal F}}_\zeta^{(i)}(x) \simeq 0$ in this region.

Combining all these contributions, one finds for the $\b -\u$ SPDs the superposition 
\begin{equation}
\widetilde{{\cal F}}_\zeta^{(i)}(x)
     = \theta(x-\zeta)\widetilde{{\cal F}}_{\zeta\,{\rm ove}}^{(i)}(x)\,+\,
       \theta(\zeta-x)\theta(x)\left[ \widetilde{{\cal F}}_{\zeta\,{\rm ann}}^{(i)}(x)
        \, +\, \widetilde{{\cal F}}_{\zeta\,{\rm res}}^{(i)}(x) \right]
                                                                      \,.    
\label{step}
\end{equation}
The relative importance of the overlap contribution to the SPDs 
on the one side and the sum of annihilation and resonance one on 
the other side, change with the momentum transfer as a consequence 
of the relation (\ref{q-z}). At large recoil, $q^2 \simeq 0$, the
annihilation and resonance parts do not contribute while they dominate 
at small recoil, $q^2\simeq q^2_{\rm max}$. The superposition 
(\ref{step}) is controlled by the skewedness parameter $\zeta$ in an 
unambiguous way, i.e.\ there is no danger of double counting. 

For the numerical estimate of the overlap contribution a simple
Gaussian \wf{} is used to describe the pion's valence Fock state
\begin{equation}
\Psi_\pi (x,\vk) = \frac{\sqrt6}{f_\pi} \, 
        \exp\left[- \frac{1}{8\pi^2 f_\pi^2} \, \frac{\vk^2}{x \,
        (1-x)}\right] \,.
\label{pi-wf}
\end{equation}
$f_\pi$ (=132\mev) is the usual pion decay constant. 
The \wf{} (\ref{pi-wf})  
has been tested against experiment and found to work satisfactorily in
many exclusive reactions involving pions (e.g.\ \cite{KR,JKR}).
It is also supported by theoretical studies, e.g.\ \cite{braun}.

For the $\b \dbar$ \wf{} of the $B$ meson a slightly modified version 
of the Bauer-Stech-Wirbel (BSW) function \cite{Wirbel:1985ji} is used
\begin{equation}
\Psi_B(x,\vk) \propto  f_B \, x \, (1-x) \, 
        \exp\left[-a_B^2 \, [M_B^2 \, \left(x - x_0\right)^2 +  \vk^2 ]\right]\,.
\label{B-wf}
\end{equation}
$m_{\b}$ is taken to be 4.8 GeV, $a_B=1.51 \gev^{-1}$, $f_B=180 \mev$ 
and the \wf{} is normalized to unity. The distribution amplitude
exhibits a pronounced peak, its position is
approximately at $x\simeq x_0=m_{\b}/M_B$. 

In principle, the overlap parts of the SPDs receive
contributions  from all Fock states. 
The generalization of the overlap representation (\ref{par-spd}) to
higher Fock states is a straightforward application of the methods
outlined in \cite{Diehl:1998kh}. Using suitably generalized
$N$-particle wave functions, one can show that the higher Fock state 
contributions to the SPD $\widetilde{{\cal F}}_{\zeta\,{\rm ove}}^{(1)}$ 
are very small and can be neglected; they represent power corrections 
$(\bar{\Lambda}/M_B)^{n(N)}$.

In order to estimate the resonance contribution the same ansatz as for
the $B$-meson is employed for the $B^*$-meson \da{}. Its explicit 
form is however irrelevant for the transition form 
factors. The product of the $B^*$ decay constant, $f_{B^*}$
and the $B B ^* \pi$ coupling constant is taken to be $20 f_B$
\cite{Khodjamirian:1998ji}.
\begin{figure}[t]
\begin{center}
\psfig{file=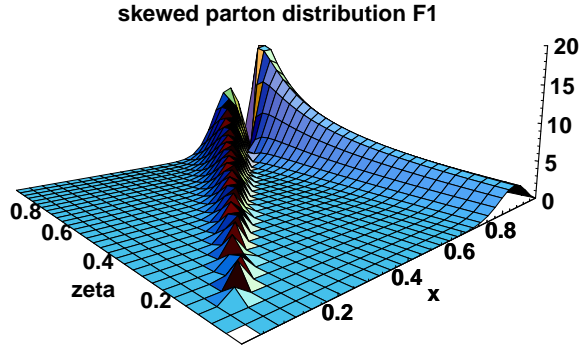, bb= 65 245 545 560, width=8cm}
\end{center}
\vspace*{-0.5cm}
\caption{The SPD $\widetilde{{\cal F}}_\zeta^{(1)} (x)$ vs. $x$
and $\zeta$.}
\label{fig2a}
\vspace*{-0.5cm}
\end{figure}
The numerical results for the $\b$-$\u$ SPD   
$\widetilde{{\cal F}}_{\zeta}^{(1)}$ are shown in Fig.\
\ref{fig2a}. Both the contributions exhibit charcteristic bumps which 
are generated by the pronounced peaks in the $B$ and $B^*$ \das.

\section{$B$-$\pi$ FORM FACTORS}
The $B - \pi$ form factors $F^{(i)}$ can be evaluated form the SPDs
through the reduction formula (\ref{fff}). Since the form factors
$F_+$ and $F_0$, being linearly related to the $F^{(i)}$, are more
suitable in applications to decay processes, I only present results
for them. 
In addition to the overlap and resonance contributions the form factors
also receive contributions from perturbative QCD where a hard gluon, 
with a virtuality of the order of $M_B^2$, is exchanged between the 
struck and the spectator quark. In Ref.\ \cite{Dahm:1995ne} the 
perturbative contributions have been evaluated at large recoil within 
the modified perturbative approach and one can make use of these results. 
At small recoil the perturbative contributions cease to be reliable
because of the small virtualities some of the internal off-shell
quarks and gluons acquire in this region. 

\begin{figure}[t]
\begin{center}
\unitlength0.8cm
\epsfclipon
{\psfig{file=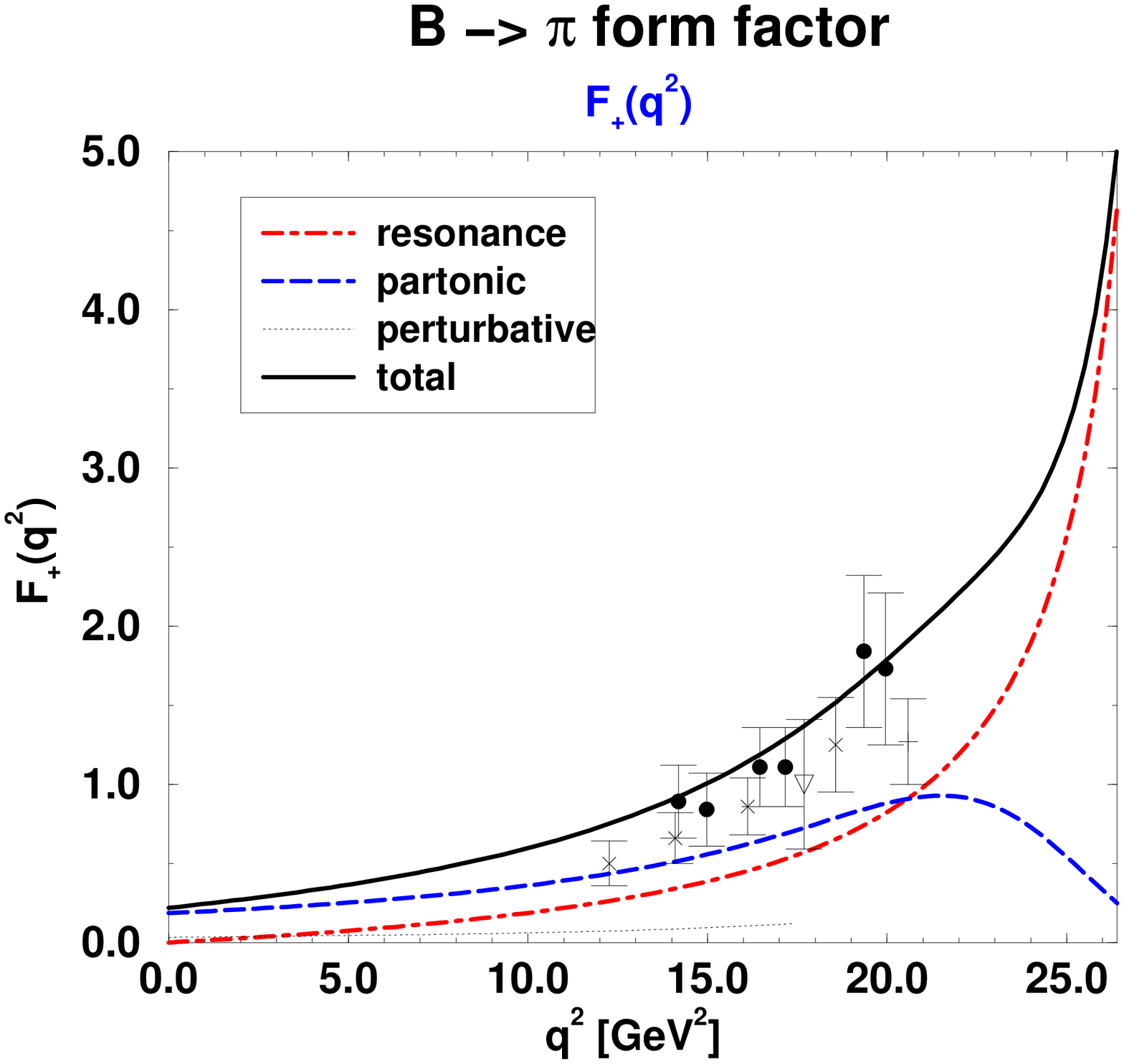, bb=60 70 570 625, 
	width=7.5\unitlength, angle=-90}}
\hskip2em
{\psfig{file=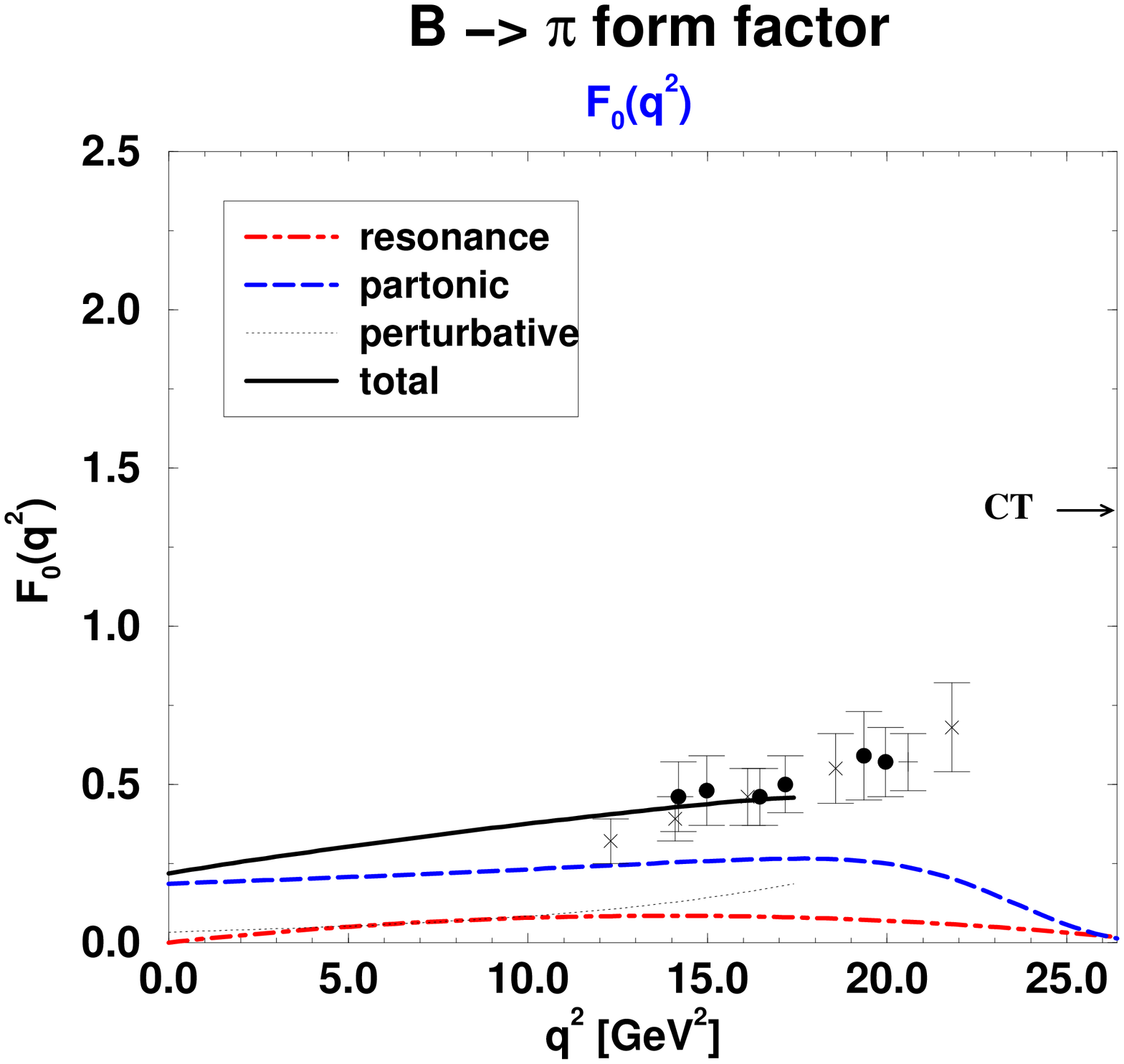, bb=60 70 570 625, 
	width=7.5\unitlength, angle=-90}}
\end{center}
\vspace*{-1.0cm}
\caption{The form factors $F_+(q^2)$ and $F_0(q^2)$ vs. momentum transfer. 
Our predictions (solid lines) for the form factors are decomposed into
resonance, overlap and perturbative contributions. The lattice QCD data, 
taken from Ref.\ \cite{Flynn:1996rc}, are shown for comparison. CT
indicates the Callan-Treiman value, $f_B/f_\pi$.} 
\vspace*{-0.5cm}
\label{fig3}
\end{figure}

Numerical results for the form factors are displayed in Fig.\ \ref{fig3}.
In the case of the form factor $F_+$ one sees the dominance of the
overlap contribution at large recoil while the resonance contribution
takes the lead at small recoil (cf.\ Eq.\ (\ref{step})). 
The perturbative contribution provides a correction to $F_+$ of the
order of $10\%$ at large recoil and can be neglected at small recoil.
The sum of the three contributions to $F_+$ is in fair 
agreement with lattice QCD results \cite{Flynn:1996rc}.
Due to the absence of the $B^*$ pole the form factor $F_0$ behaves
differently; it is rather flat over the entire range of momentum
transfer. The perturbative contribution makes up a substantial
fraction of the total result for $F_0$ at intermediate momentum
transfer. Since, as is mentioned above, 
it becomes unreliable for $q^2 \gsim 18 \gev^2$ $F_0$ cannot reliably
be predicted at large $q^2$. A calculation of $F_0$ in
that region would also require a
detailed investigation of the scalar $B\pi$ resonances of which not
much is known at present. Despite of this drawback 
the  results for this form factor are also in fair agreement
with the lattice QCD results \cite{Flynn:1996rc} and, in tendency, seem 
to extrapolate to the $B$-sector analogue of the Callan-Treiman value.  

An assessment of the theoretical uncertainties of the predictions for
the $B\to\pi$ transition form factors leads to an uncertainty of about
$20 - 25 \%$. It includes estimates of: Sudakov suppressions in the 
end-point region ($x\to 1$), contributions from two-particle 
twist-three \wfs{}s, deviations from the asymptotic form of the pion \da{},
uncertainties of the input parameters ($g_{BB^*\pi}$,$f_B$, $f_{B^*}$, 
$m_{\b}$) and from the order $\bar{\Lambda}/M_B$ corrections, 
In particular for the form factors at maximum
recoil, $F_+(0)= F_0(0)$ a value of $0.22\pm 0.05$ is obtained. 

With the form factors at hand one can evaluate the semi-leptonic decay
rates $\bar{B}^0\to \pi^+ \ell^- \bar{\nu}_\ell$. 
For the branching ratio of the light-lepton modes one finds
\be
{\cal B}[\bar{B}^0\to \pi^+ e \bar{\nu}_e] \simeq {\cal B}[\bar{B}^0 \to \pi^+
\mu \bar{\nu}_\mu] = 1.9 \cdot 10^{-4} 
\cdot \left(\frac{|V_{\u\b}|}{0.0035}\right)^2 .
\end{equation} 
The theoretical uncertainty of this prediction, dominated by that of
the overlap contribution, amounts to about
$30\%$. This result is to be compared with the CLEO measurement
\cite{ale:96}: $(1.8\pm0.4\pm0.3\pm0.2) \cdot 10^{-4}$. 
For the $\tau$ channel one obtains a value of $1.5\cdot 10^{-4}$ for
the branching ratio ($|V_{\u\b}|=0.0035$).

The exclusive $B$-decays into pairs of pions are usually calculated on the
basis of a factorization hypothesis according to which the decay
amplitudes can be written as a product of two weak current matrix
elements
\be
{\cal M}\,=\, \frac{G_F}{\sqrt{2}}\, V^*_{\u\d} V_{\u\b}\,
                     \langle \pi^-;q|J^\mu_W| 0\rangle
                 \langle \pi^+; p' |J_\mu^W| \overline{B}^0; p\rangle\,.
\ee
The first matrix element defines the usual pion decay constant
($\propto f_\pi q^\mu$) while the second one defines the $B\to\pi$
transition form factors (\ref{current}). The
factorizing contribution alone leads to the following branching ratio
\be
{\cal B}(\bar{B}^0\to\pi^+\pi^-) = 10.5\cdot 10^{-6} \,
             \left(\frac{|V_{\u\b}|}{0.0035}\right)^2
             \, \left | \frac{F_+(0)}{0.33}\right |^2.
\ee 
Ignoring the short-distance corrections which seem to amount to about 
$10 -20\%$ \cite{ali:1998}, and choosing $|V_{\u\b}|=0.0035$, one 
finds agreement between the prediction for $F_+(0)$ from the SPD
approach and the recent CLEO measurement \cite{cronin} for the 
$\bar{B }^0\to\pi^+\pi^-$ branching ratio of $(4.3\, {}^{+1.6}_{-1.4} 
                                     \pm 0.5)\cdot 10^{-6}$. 
The experimental value is much smaller than expected (based
on $F_+(0)\simeq 0.3 - 0.33$) and a revision of the theoretical
analysis of exclusive $B$-decays seems to be required.
\section{CONCLUSIONS}
The $\b$-$\u$ SPDs are calculated from light-cone \wf{}
overlaps and a contribution from  the $B^*$ resonance. The
chief advantage of the SPD approach is that the skewedness parameter
clearly separates the overlap from the resonance contribution and
both the contributions can be added unambigously. The $B\to\pi$ 
transition form factors are calculated from the $\b -\u$ SPDs by means of
reduction formulas. $F_+$ is obtained in the entire range of momentum
transfer and $F_0$ up to about 18\gev$^2$. In particular, a value of
$0.22\pm 0.05$ is found for the form factors at maximum recoil. This value
appears to be in agreement with the recent CLEO measurement
\cite{cronin} of $B\to \pi\pi$ decays (if the latter process is
analysed on the basis of the factorization hypothesis).
The prediction for the total decay for the process 
$\bar{B}^0\to\pi^+e\bar{\nu}_e$ is also in agreement with a CLEO 
measurement \cite{ale:96}. In both the cases, the $\pi\pi$ and the
semi-leptonic decay, a value of 0.0035 is used for the CKM matrix 
element $V_{\u\b}$.

\end{document}